\documentstyle[12pt]{article}

\begin{document}

\newcommand{\Om}{\Omega}
\newcommand{\df}{\stackrel{\rm def}{=}}
\newcommand{\co}{{\scriptstyle \circ}}
\newcommand{\de}{\delta}
\newcommand{\lb}{\lbrack}
\newcommand{\rb}{\rbrack}
\newcommand{\rn}[1]{\romannumeral #1}
\newcommand{\msc}[1]{\mbox{\scriptsize #1}}
\newcommand{\dsp}{\displaystyle}
\newcommand{\scs}[1]{{\scriptstyle #1}}

\newcommand{\ket}[1]{| #1 \rangle}
\newcommand{\bra}[1]{| #1 \langle}
\newcommand{\vac}{| \mbox{vac} \rangle }

\newcommand{\e}{\mbox{{\bf e}}}
\newcommand{\va}{\mbox{{\bf a}}}
\newcommand{\bc}{\mbox{{\bf C}}}
\newcommand{\br}{\mbox{{\bf R}}}
\newcommand{\bz}{\mbox{{\bf Z}}}
\newcommand{\bq}{\mbox{{\bf Q}}}
\newcommand{\bn}{\mbox{{\bf N}}}
\newcommand {\eqn}[1]{(\ref{#1})}

\newcommand{\cp}{\mbox{{\bf P}}^1}
\newcommand{\n}{\mbox{{\bf n}}}
\newcommand{\sbz}{\msc{{\bf Z}}}
\newcommand{\sn}{\msc{{\bf n}}}

\newcommand{\be}{\begin{equation}}\newcommand{\ee}{\end{equation}}
\newcommand{\bea}{\begin{eqnarray}} \newcommand{\eea}{\end{eqnarray}}
\newcommand{\ba}[1]{\begin{array}{#1}} \newcommand{\ea}{\end{array}}

\newcommand{\cleqn}{\setcounter{equation}{0}}
\makeatletter
\@addtoreset{equation}{section}
\def\theequation{\thesection.\arabic{equation}}
\makeatother

\def\np{Nucl. Phys. {\bf B}}
\def\pl{Phys. Lett. {\bf B}}
\def\mpl{Mod. Phys. {\bf A}}
\def\ijmp{Int. J. Mod. Phys. {\bf A}}
\def\cmp{Comm. Math. Phys.}
\def\prd{Phys. Rev. {\bf D}}

\def\vu{\vec u}
\def\vs{\vec s}
\def\vv{\vec v}
\def\vt{\vec t}
\def\vn{\vec n}
\def\ve{\vec e}
\def\vp{\vec p}
\def\vk{\vec k}
\def\vx{\vec x}
\def\vz{\vec z}
\def\vy{\vec y}


\begin{flushright}
La Plata Th-00/07\\November 17, 2000
\end{flushright}

\bigskip

\begin{center}

{\Large\bf Correlation functions in the non-commutative Wess-Zumino-Witten model}
\footnote{ This work was partially supported
by CONICET, Argentina}

\bigskip
\bigskip

{\it \large Adri\'{a}n R. Lugo}
\footnote{
{\sf lugo@dartagnan.fisica.unlp.edu.ar}
}
\bigskip

{\it Departamento de F\'\i sica, Facultad de Ciencias Exactas \\
Universidad Nacional de La Plata\\ C.C. 67, (1900) La Plata,
Argentina}
\bigskip
\bigskip

\end{center}

\begin{abstract}
We develop a systematic perturbative expansion and compute the one-loop two-points,
three-points and four-points correlation functions in a non-commutative version of the
$U(N)$ Wess-Zumino-Witten model in different regimes of the $\theta$-parameter showing
in the first case a kind of phase transition around the value
$\theta_c = \frac{\sqrt{p^2 + 4\,m^2}}{\Lambda^2\,p}$, where $\Lambda$ is a ultraviolet cut-off
in a Schwinger regularization scheme.
As a by-product we obtain the functions of the renormalization group, showing they are
essentially the same as in the commutative case but applied to the whole $U(N)$ fields;
in particular there exists a critical point where they are null, in agreement with a recent
background field computation of the beta-function, and the anomalous dimension of the Lie
algebra-valued field operator agrees with the current algebra prediction.
The non-renormalization of the level $k$ is explicitly verified from the four-points
correlator, where a left-right non-invariant counter-term is needed to
render finite the theory, that it is however null on-shell.
These results give support to the equivalence of this model with the commutative one.

\end{abstract}

\bigskip

\section{Generalities}
\cleqn
Non commutative (NC) field theories have recently attracted attention because of its
comparison as effective theories in the context of D-brane physics \cite{seiwit}.
Due to this fact mostly studied were NC gauge theories \cite{cds}, \cite{luis} as well as
toy models of scalar non derivative field theories \cite{minraamsei}, \cite{gomis2}.
They are generally formulated as extensions of ordinary field theories where the usual
point-to point product is replaced by the ``*" product; only in two dimensions this
procedure preserves Lorentz (or $SO(2)$ in euclidean formulations) invariance, however
it necessary breaks (if there were) scale invariance explicitly.
Being two dimensional conformal field theories (CFT) a subject vastly studied during the last
two decades due to its implications in Statistical Mechanics and Solid State Physics other
than in the perturbative formulation of Superstring Theories, a question naturelly arise:
do the NC extensions of two dimensional CFT define at quantum level another CFT?
And if so, what?
In the case of free field theories (free bosons and fermions) the answer is yes, but in a
trivial way because both theories indeed coincide explicitely.
And among the interacting theories more or less tractable by current algebra methods are
the Wess-Zumino-Witten (WZW) models \cite{wit}.
It is the aim of this paper to study  perturbative aspects of these peculiar models that
include infinite interaction vertices containing derivatives.
For conventions adopted about NC spaces, groups definitions, etc., we refer the reader to
Appendix A.

For definiteness we consider the non-commutative version of the $U(N)$ WZW model in
euclidean space defined by the bare action
\bea
S[g] &=& \frac{1}{\lambda_0{}^2}\; \left( I_0 [g] + I_{WZ}[\tilde g] \right)\cr
I_0 [g] &=& \frac{1}{2}\;\int_{\Sigma} d^2\vx \; Tr\left( L_i(g) L_i(g)\right)\cr
I_{WZ}[\tilde g] &=& i\, \frac{\alpha_0}{3}\;\int_{B}
Tr\,\left( L(\tilde g)\wedge L(\tilde g)\wedge L(\tilde g)\right)\label{accion}
\eea
where $\lambda_0$ is the coupling constant of the theory and $\alpha_0$ will be eventually
identified with $\frac{\lambda_0^2\, k_0}{4\pi}$ with $k_0$ a parameter not to be renormalized.
\footnote{
This fact is enforced in the commutative case where by topological reasons it must be an
integer; here we will just verify that this property continues to hold in the NC case at least
perturbatively, however the integer character is certainly not yielded by our analysis,
see \cite{dabro} in relation to this topic.
}
The three-dimensional manifold ``B" is taken to be a cylinder with the top and bottom
disks of infinite radius (two $\Re^2$'s) parametrized by $\vx= (x^1, x^2)$ while the height
variable $s\in\Re $.
The boundary conditions on $\tilde g (\vx, s)$ are
\bea
\tilde g (\vx, s)&\stackrel{x\rightarrow\infty}{\longrightarrow}& 1\cr
\tilde g (\vx, s)&\stackrel{s\rightarrow -\infty}{\longrightarrow}& 1\cr
\tilde g (\vx, s)&\stackrel{s\rightarrow\infty}{\longrightarrow}& g(\vx)\label{bc}
\eea
Unless specified the contrary all the field products along the paper are understood as
$``*"$ products with parameter $\theta^{ij} \equiv \theta\;\epsilon^{ij},\; \theta^{is} = 0$,
where the matrix $\epsilon = i\,\sigma_2$ as usual in two dimensions and $\theta>0$.
Under these conditions usual properties of the commutative case hold and the effective
degrees of freedom are represented by $g(\vx)$; in particular the Polyakov-Wiegmann formula
\bea
S[g\, h] &=& S[g] +  S[h] + \frac{1}{\lambda^2}\;\int d^2\vx\;
Tr\left(\,P_{-}^{ij}\; L_i(g) \; R_j(h)\right)\cr
P_{\pm}^{ij} &\equiv& \delta^{ij} \pm i\; \alpha_0\;\epsilon^{ij}
\label{polwieg}
\eea
as well as cyclic properties under traces are valid.
The first order variation of the action is ($z=x^1 + i\,x^2$)
\bea
\delta S[g] &\equiv& S[g+\delta g] - S[g] =
-\frac{1}{\lambda_0{}^2}\;\int\;d^2\vx\;P_+^{ij}\;Tr\left(g^{-1}\,\delta g\;
\partial_i\,L_j(g)\right)\cr
&=& \frac{1}{\lambda_0{}^2}\int \, i\, d\overline z \wedge dz\;
Tr\,\left( g^{-1}\; \delta g\; \left( 2\, g^{-1}\, \partial_z\partial_{\overline z} g
+ (1+\alpha_0)\, \partial_z g^{-1}\; \partial_{\overline z} g\right.\right. \cr
&+& \left.\left.(1-\alpha_0)\, \partial_{\overline z} g^{-1}\; \partial_z g\right)\right)
\label{varac}
\eea
In the critical case where (for definiteness) $\alpha_0 = 1$ yields as equations of motion
the usual conservation of the currents
$J\equiv \frac{k}{4\pi}\, R_z\;\; ,\; \overline J \equiv -\frac{k}{4\pi}\,L_{\overline z}$
\be
\partial_{\overline z}\, J = \partial_z \,{\overline J} = 0
\ee
The action (\ref{accion}) is invariant under global transformations
$ g \rightarrow h_{L}g h_{R}$, with $h_{L} (h_{R})$ belonging to any subgroup $H_{R}
(H_{L})$ of $U(N)$ isomorphic to $U(p), p\leq N,$.
In the commutative critical case conformal invariance raises this invariance to holomorphic and
anti-holomorphic dependence, being the generators of these transformations the momenta of the
currents $J, \overline J$ that satisfy the standard level $k$ left and right Kac-Moody algebra.
In the NC case we have not certainly conformal invariance due to the introduction of the
``*" product; however it seems that the holomorphic character of the currents (now defined
with the ``*" product) continues to hold.
However we are not in conditions of asserting that they verify a current algebra because we
have not to our disposal a hamiltonian formulation which yielded a canonical quantization of
the theory, the main obstacle being the infinite time derivatives in the lagrangean
that give a non-local character to the theory.
\footnote{
We have no clear at all that the method advocated in reference \cite{wit} based in a Poisson
structure determined by equation (\ref{varac}) led to a right answer in the NC context due to
the presence of the ``*" product, other the fact we do not know of generalizations of it.
Recent work in reference \cite{gomis1} could help to address this question that certainly
deserves further study.
}
Then the existence or not of a conformal structure is in our opinion an open question, being
one of the goals of the present work to shield some light about it.

\section{The perturbative expansion}
\cleqn

In reference \cite{fur} a computation of the one-loop beta function was made by using the
background fiel method.
It is known however that this method to compute the OPI effective action
$\Gamma[\Phi]$ is not convenient beyond one-loop; from its path integral expression
\bea
\Delta\Gamma[\Phi] &\equiv& \Gamma[\Phi] - S[\Phi]= - \ln\; \int\, D\xi\; e^{- S_q[\xi;\Phi] +
\int\, d^d \vx\, \frac{\delta\Delta\Gamma[\Phi]}{\delta\Phi}\; \xi}\cr
S_q[\xi;\Phi] &\equiv& S[\Phi + \xi] - S[\Phi] -\int\, d^d \vx\,\frac{\delta
S[\Phi]}{\delta\Phi}\;\xi\label{efac}
\eea
we see that at one-loop the source term is not present and the computation is straight,
but beyond one-loop it must be included in the form of $\vx$-dependent one-point vertices,
the procedure becoming more and more involved.
This drawback is common to any QFT but in theories with fields living on group manifolds
(as the present case) another one is present and has to do with the fact that if we write
the group-valued field $g=\exp(\phi)$ with $\phi$ the fields of the theory living in the Lie algebra
then the splitting  corresponding to (\ref{efac}) in a background and a quantum part $\phi = \Phi + \xi$
identifying $g_0\equiv\exp(\Phi)$ as the background field living in the group yields an awful expansion;
the best suited splitting used in references \cite{fur}, \cite{boss} is $g = g_0 \; \exp(\pi)$ but in order
to be used beyond one-loop through (\ref{efac}) we would need the NC Baker-Campbell-Haussdorf formula to get
$\xi = \xi[\pi;g_0]$ through the equality
\be
g =\exp(\Phi + \xi) \leftrightarrow   \exp(\pi) = \exp(-\Phi) \; \exp(\Phi +\xi)
\ee
and we must face another hard problem.
So we prefer to develop a systematic perturbative series in the standard way.

To regulate in the infrared we add to (\ref{accion}) a mass term \cite{boss}
\be
S_m [g] = \frac{m_0^2}{2\,\lambda_0^2}\;\int d^2\vx \; Tr\left( g + g^{-1} - 2 \right)
\label{masa}
\ee
As showed in \cite{boss} dimensional regularization could be consistently used to regulate
in the ultraviolet; however we need to add extra-dimensions terms, a new coupling, etc;
to evite these drawbacks we adopt a Schwinger-like prescription \cite{zinn} and take the free
propagator
\be
\frac{1}{k^2 + m_0{}^2}\;\longrightarrow\;
\tilde G(k^2) =\frac{e^{-\frac{k^2}{2\Lambda^2}}}{k^2 + m_0{}^2}
\ee
where $\Lambda$ is the UV cut-off.


The group element is parametrized as $g\equiv \exp\left( \lambda_0 \pi_0\right)$ where
$\pi_0(\vx) = \pi_0^a (\vx)\, X_a$ is the quantum field living on $u(N)$ and obeying from
(\ref{bc}) the boundary condition $\pi_0(\vx)\stackrel{x\rightarrow\infty}{\longrightarrow}0$.
The action is then written
\bea
S[g] &=& S_q[\pi_0] + \sum_{n\geq 3} S^{n}[\pi_0]\cr
S_q[\pi_0] &=& \frac{1}{2}\; \int \; \frac{d^2 \vk}{(2\,\pi)^2}\;
\kappa_{b_1b_2}\;\tilde G(k)^{-1}\;
\tilde\pi_0^{b_1}(\vk)\;\tilde\pi_0^{b_2}(-\vk)\label{actionpert}
\eea
with $S^n[\pi_0]$ the vertices given in Appendix A.
Here we quote explicitly in momentum space the first four ones including $S_m$.
By denoting $\vk\times\vp \equiv \epsilon^{ij}\, k_i\,p_j$ we have
\bea
S^3[\pi_0] &=& \prod_{l=1}^{3}\; \left(\int \frac{d^2\vk_l}{(2\pi)^2}\; \tilde\pi_0^{b_l}(\vk_l)
\right) (2\pi)^2\,\delta^2\left(\sum_{l=1}^{3} \vk_l\right)\; \frac{\alpha_0\,\lambda_0}{i\,9}\;
\kappa_{b_1b_2b_3}\; v_c^3(\vk_1,\vk_2,\vk_3)\cr
v_c^3(\vk_1,\vk_2,\vk_3) &=& \exp\left(\frac{\theta}{2i}\; \vk_1\times\vk_2\right)\;
\vk_1\times\vk_2 + (cyclic)
\eea
\bea
S^4[\pi_0] &=& \prod_{l=1}^{4}\; \left(\int \frac{d^2\vk_l}{(2\pi)^2}\; \tilde\pi_0^{b_l}(\vk_l)
\right) (2\pi)^2\,\delta^2\left(\sum_{l=1}^{4} \vk_l\right)\; \frac{\lambda_0^2}{4!\,2}\;
\kappa_{b_1\dots b_4}\; v_c^4(\vk_1,\dots,\vk_4)\cr
v_c^4(\vk_1,\dots ,\vk_4) &=& \exp\left(\frac{\theta}{2i}\;
(\vk_1\times\vk_2 + \vk_3\times\vk_4 )\right)\; \left( (\vk_1-\vk_2)\cdot\vk_3+ \frac{m_0{}^2}{2}
\right)\cr
&+& (cyclic)
\eea
\bea
S^5[\pi_0] &=& \prod_{l=1}^{5}\; \left(\int \frac{d^2\vk_l}{(2\pi)^2}\; \tilde\pi_0^{b_l}(\vk_l)
\right) (2\pi)^2\,\delta^2\left(\sum_{l=1}^{5} \vk_l\right)\;
\frac{\alpha_0\,\lambda_0{}^3}{i\,300}\;
\kappa_{b_1\dots b_5}\; v_c^5(\vk_1,\dots,\vk_5)\cr
v_c^5(\vk_1,\dots,\vk_5) &=& \exp\left(\frac{\theta}{2i}\;
( \vk_1\times\vk_2 + \vk_1\times\vk_3+\vk_2\times\vk_3+\vk_4\times\vk_5 )\right)\;
\vk_1\times(\vk_2 + 3\,\vk_4)\cr
&+& (cyclic)
\eea
\bea
S^6[\pi_0] &=& \prod_{l=1}^{6}\; \left(\int \frac{d^2\vk_l}{(2\pi)^2}\; \tilde\pi_0^{b_l}(\vk_l)
\right) (2\pi)^2\,\delta^2\left(\sum_{l=1}^{6} \vk_l\right)\; \frac{\lambda_0^4}{6!\,3}\;
\kappa_{b_1\dots b_6}\; v_c^6(\vk_1,\dots,\vk_6)\cr
v_c^6(\vk_1,\dots ,\vk_6) &=&
\exp\left(\frac{\theta}{2\, i}\;\left(
 \vk_1\times\vk_2 + \vk_1\times\vk_3+\vk_2\times\vk_3+\vk_4\times\vk_5
+\vk_4\times\vk_6+\vk_5\times\vk_6 \right)\right)\cr
& & \times\left( \vk_6\cdot(-3\,\vk_3 + 4\,\vk_4 - \vk_5) + \frac{m_0{}^2}{2} \right)
+ (cyclic) \label{vermom}
\eea
We have expressed them in terms of vertex functions $v_c^n$ cyclic-invariant in momenta
because this form is very useful in the evaluation of Feynmann diagrams.
Also the momentum conservation is taken into account to simplify them.
\footnote{
Taking into account this constraint the exponential $\theta$-dependent factors result cyclic
invariant.
}
We remember finally that loop corrections correspond to powers of $\lambda^2$ (see
Section 5 for definitions of the renormalized parameters).

\section{The one-loop two-points function}
\cleqn
In what follows we write a generic OPI correlation function in momentum space as
\be
\Gamma\left(\vk_1,\dots,\vk_n\right) = (2\pi)^2\, \delta^2 \left(\sum_{i=1}^n\,\vk_i\right)\;
\tilde\Gamma\left(\vk_1,\dots,\vk_n\right)
\ee
The tree level  OPI two-point function
\footnote{
Most precisely the inverse of the connected two-point function; it is OPI for vertex
functions, see for example \cite{zinn}.
}
is
\be
\tilde\Gamma_{a_1a_2}^{(0)} (p) = \kappa_{a_1a_2}\, ( p^2 + m^2)
\ee
At one-loop the contribution to the OPI two-point function consist of two parts that can be
visualized as standard graphs; the first one labeled $``(a)"$ contains two three-point vertices
and is given by
\footnote{
Indeed there exist a third contribution of this kind, a ``blob", that is identically null.
}
\be
\tilde\Gamma_{a_1a_2}^{(1a)}(p) = \alpha^2\,\lambda^2\; \left(
N\, \kappa_{a_1a_2}\; f(p;0) + \kappa_{a_1}\,\kappa_{a_2}\; f(p;\theta)\right)\label{unoa}
\ee
We note the second term in (\ref{unoa}) is like a ``non planar" diagram in the language of
\cite{minraamsei}.
The function $f(p;\theta)$ is exactly computed in Appendix B in various regimes of the
parameter $\theta$; its large $\Lambda$ result is
\be
f(p; 0) = -\frac{p^2}{8\pi}\;\left( Ei( -\frac{m^2}{\Lambda^2} -\frac{p^2}{4\Lambda^2})
- 1 + \sqrt{1+ \frac{4m^2}{p^2}}\; \ln\frac{4\,m^2}{p^2 + 4\,m^2} +
\ln\frac{\sqrt{p^2 + 4\,m^2} + p}{\sqrt{p^2 + 4\,m^2} - p}\right)\label{efecero}
\ee
and if $\theta\neq 0$ (${\tilde p}^2\equiv p\,\sqrt{p^2 + 4\,m^2}$),
\bea
f(p;\theta) &=& - \frac{{\tilde p}^2}{8\pi}\; \left(
e^{\frac{\theta\,{\tilde p}^2}{2}}\; Ei\left(\frac{\theta\,{\tilde p}^2}{2}\right)
+ e^{-\frac{\theta\,{\tilde p}^2}{2}}\; Ei\left(-\frac{\theta\, {\tilde p}^2}{2}\right)
+ f_m(p;\theta)\right)\cr
f_m(p;\theta) &=& \frac{4}{\theta\,{\tilde p}^2}\;\int_0^{\theta\,m\,p}\; dx\;
\frac{x\; (x\,K_0(x))'}{\sqrt{\frac{\theta^2\,{\tilde p}^4}{4}-x^2}}\;\;
\stackrel{m\rightarrow0}{\longrightarrow}0\label{efeteta}
\eea
where $K_0$ is a Bessel function.

The second contribution labeled $``(b)"$ comes from a diagram containing a quartic vertex with
a self-contraction and is given by
\bea
\tilde\Gamma_{a_1a_2}^{(1b)}(p) &=& -\frac{\lambda^2}{6}\; \left(
N\,\kappa_{a_1a_2}\; \left( g_1(p;0) + (p^2 + 2m^2) \; g_2(p;0)\right)\right.\cr
&+&\left.\kappa_{a_1}\,\kappa_{a_2}\;\left( g_1(p;\theta) + (p^2 - m^2) \; g_2(p;\theta)\right)
\right)\cr
g_i(p;\theta) &=& \int\;\frac{d^2\vk}{4\pi^2}\;{\tilde G}(k)\;  e^{i\,\theta\,\vp\times
\vk}\;\times\; \left\{\begin{array}{ll}k^2 &\;\;,\;\; i=1\\1&\;\;,\;\; i=2\end{array} \right.
\label{unob}
\eea
The functions $g_i$ are easily computed as we made with $f(p;\theta)$ in Appendix B; for
large $\Lambda$ we have
\bea
g_1(p;\theta) &=&\left\{\begin{array}{ll} \frac{1}{4\,\pi}\left( 2\,\Lambda^2 + m^2\;
Ei(-\frac{m^2}{2\Lambda^2})\right)&\;\;,\;\;\theta\,p=0\\
\frac{1}{\theta^2}\; \delta^2(\vp)-\frac{m^2}{2\,\pi}\; K_0(\theta\,m\,p)&\;\;,\;\;
\theta\, p\neq 0\end{array}\right.\cr
g_2(p;\theta) &=& \left\{\begin{array}{rr}
-\frac{1}{4\,\pi}\; Ei(-\frac{m^2}{2\Lambda^2})&\;\;,\;\;\theta\, p=0\\
\frac{1}{2\,\pi}\; K_0(\theta\,m\,p)&\;\;,\;\;\theta\, p\neq 0\end{array}\right.\label{g}
\eea
Some remarks are in order.
In first place we note that as remarked in Appendix B, $f(0;\theta)=0$ and the limits
$\Lambda\rightarrow\infty$ and $p=0$ commute; this is due to the factorization of the
external momentum $p$ coming from the vertex.
However it is not the case for the tadpole diagram where we see from (\ref{g}) that
the presence of the scale $\theta$ induces, at fixed $\Lambda$, standard logarithmic
singularities in $p=0$ and a $\delta$-type singularity;  similar term was recently noted in
\cite{gomis2} in the context of a $2+1$ dimensional non-relativistic non-commutative field
theory; an analogous term is shown to be present in the relativistic theory in a certain limit.
It is also remarked there that it is not possible to apply the usual normal ordering
prescription of commutative field theories to set the tadpole to zero because of its dependence
on the external momentum coming from the non-planar diagrams.
However they will be irrelevant to the computation of the renormalization group functions
as will be shown in detail in Section $5$.
In second place, we should hope in the commutative case to have only contributions proportional
to the traceless part corresponding to $SU(N)$ because the $U(1)$ trace part is just a free
field, i.e. the correlation function should be proportional to the tensor
\be
\kappa_{a_1a_2}^{(tr)} = \kappa_{a_1a_2} +
\frac{\kappa_{a_1}\;\kappa_{a_2}}{N}\;\;\;,\;\;\kappa^{a_1a_2}\, \kappa_{a_1a_2}^{(tr)}=0
\ee
This is evident in (\ref{unoa}), not so in (\ref{unob}), however it becomes true in the
massless limit.
The explanation of this fact lies in the presence of the mass term (\ref{masa}) that
introduces interactions for the trace part not present in the classical theory.

\section{The one-loop three-points function}

From (\ref{vermom}) we read that the OPI three-point function at tree level is given by
\be
\tilde\Gamma_{a_1a_2a_3}^{(0)}(\vp) =
-\frac{i\,\alpha\,\lambda}{3}\;\kappa_{a_1a_2a_3}\;\vp_1\times\vp_2\;
e^{-i\frac{\theta}{2}\,\vp_1\times\vp_2} + (perm.)
\ee
where ``(perm.)" implies all permutation terms of external legs.
Standard  series expansion tells us that at one loop there exist three contributions.
The first labeled ``$(a)$" is a tadpole with a five-points vertex insertion,
\be
\tilde\Gamma_{a_1a_2a_3}^{(1a)}(\vp) =
\frac{i\,\alpha\,\lambda^3}{12}\;\vp_1\times\vp_2\;e^{-i\frac{\theta}{2}\,\vp_1\times\vp_2}\;
\left(N\,\kappa_{a_1a_2a_3}\; g_2(p_3,0) \; + \kappa_{a_1a_2}\kappa_{a_3}\;g_2(p_3;\theta)
\right) + (perm.)\label{31a}
\ee
where $g_2(p;\theta)$ is given in (\ref{g}).
The second contribution labeled ``$(b)$" contains a three-points vertex and a four-points
vertex and is given by
\bea
\tilde\Gamma_{a_1a_2a_3}^{(1b)}(\vp) &=&
\frac{i\,\alpha\,\lambda^3}{12}\;e^{-i\frac{\theta}{2}\,\vp_1\times\vp_2}\;\int\,
\frac{d^2\vk}{4\,\pi^2}\; \frac{e^{-\frac{1}{2\Lambda^2}( \vk^2 + (\vk - \vp_1)^2)}}
{(\vk^2 + m^2)\,((\vk -\vp_1)^2 + m^2)}\; \vp_1\times\vk\;\cr
&\times& \left( N\,\kappa_{a_1a_2a_3}\; P_1(\vk;\vp) +
\kappa_{a_1}\,\kappa_{a_2a_3}\; e^{-i\theta\,\vp_1\times\vk}\;P_1(\vk;\vp)\right.\cr
&+& \left.\kappa_{a_2}\,\kappa_{a_3a_1}\; e^{-i\theta\,\vp_2\times\vk}\;P_2(\vk;\vp) +
\kappa_{a_3}\,\kappa_{a_1a_2}\; e^{-i\theta\,\vp_3\times\vk}\;P_2(\vk;\vp)\right)\cr
&+& (perm.)
\eea
where
\bea
P_1(\vk;\vp) &=& \vk^2 + 2\,\vk\cdot(\vp_1 +3\,\vp_3) + \vp_3{}^2 + 2\,\vp_1\cdot\vp_2
+ 2\,m^2\cr
P_2(\vk;\vp) &=& \vk^2 - \vk\cdot\vp_1 - \frac{\vp_1{}^2}{2} - \vp_2\cdot\vp_3 - m^2
\eea
Finally the third contribution labeled ``$(c)$'' contains three three-point vertices,
\bea
\tilde\Gamma_{a_1a_2a_3}^{(1c)}(\vp) &=&
-\frac{i\,(\alpha\,\lambda)^3}{3}\; e^{-i\,\frac{\theta}{2}\,\vp_1\times\vp_2}\;
\left( N\,\kappa_{a_1a_2a_3}\, F(\vp;0) + 3\,\kappa_{a_1}\,\kappa_{a_2a_3}\,
F(\vp;\theta)\right)\cr
&+& (perm.)\cr
F(\vp;\theta) &=&
\int\, \frac{d^2\vk}{4\pi^2}\; \frac{e^{-\frac{1}{2\Lambda^2}\left( \vk^2 + (\vk +
\vp_1)^2 + (\vk-\vp_2)^2 \right)+ i\,\theta\,\vp_1\times\vk}}{(\vk^2 +m^2)\,
((\vk + \vp_1)^2+ m^2)\,((\vk-\vp_2)^2 + m^2)}\cr
&\times&\; \vp_1\times\vk\;\;\vp_2\times\vk\;\;\vp_3\times(\vk + \vp_1)
\eea
The integrals are straightforwardly evaluated as in Appendix B; in order not to be
reiterative we present the large $\Lambda$ result of $(b)$ useful in the next section
\be
\tilde\Gamma_{a_1a_2a_3}^{(1b)}(\vp) =
-\frac{3\,N\,\lambda^2}{16\,\pi}\;\ln\frac{m^2}{\Lambda^2}\;
\tilde\Gamma_{a_1a_2a_3}^{(0)}(\vp)\label{31b}
\ee
In what contribution $(c)$ concerns we just say that is IR and UV {\it finite}, the reason
behind being the derivative vertices present and the factorization of powers of external
momenta respectively.

\section{Renormalization}

As a by-product of the computations made  we will obtain here the functions of the
renormalization group working in the Callan-Symanzik context \cite{pes}.
In order to get rid of the divergences we introduce renormalized fields and constants that
will define the counter-terms in the way
\bea
\pi_0(\vx) &\equiv& Z^{\frac{1}{2}}\;\pi(\vx)\cr
m_0{}^2 &\equiv& Z^{-1}\,Z_m\; m^2\cr
\lambda_0 &\equiv& Z^{-1}\;Z_\lambda{}^{\frac{1}{2}}\; \lambda\cr
\alpha_0 &\equiv& Z^{-\frac{1}{2}}\; Z_\lambda{}^{-\frac{1}{2}}\; Z_\alpha\;\alpha\label{rencons}
\eea
Let us assume $k$ is not renormalized; in the next section we will prove it.
Then the relation
\be
Z_\lambda = Z\;Z_\alpha{}^{\frac{2}{3}}\label{Zalfa}
\ee
should hold.
So hopefully three renormalization constants $Z$, $Z_\lambda$ and $Z_m$ will be enough to
make finite the theory.
All renormalized quantities will depend on a scale of renormalization $\mu$ at which they
will be defined.
Let us introduce therefore the renormalization group functions
\bea
\beta &\equiv& \mu\, \frac{d\lambda}{d\mu}\cr
\gamma_m &\equiv& \mu\, \frac{d \ln m^2}{d\mu}\cr
\gamma_\alpha &\equiv& \mu\, \frac{d \alpha}{d\mu}
\label{renfunc}
\eea
By imposing as usual the independence of $\mu$ of the bare parameters we get from
(\ref{rencons}), (\ref{renfunc}) the algebraic relation
\be
\gamma_\alpha = \frac{2\,\alpha}{\lambda}\;\beta\label{zlambda}
\ee
together with the system of equations
\bea
\left(\lambda \frac{d\ln(Z_m/Z)}{d\lambda} + 2\,\alpha \frac{d\ln(Z_m/Z)}{d\alpha}\right)\;
\frac{\beta}{\lambda} +
\left( 1 + m^2\,\frac{d\ln(Z_m/Z)}{dm^2}\right)\;\gamma_m &=&
\mu\,\frac{d\ln(Z/Z_m)}{d\mu}\cr
\left(1 + \lambda \frac{d\ln(Z_\lambda{}^\frac{1}{2}/Z)}{d\lambda} + 2\,\alpha
\frac{d\ln(Z_\lambda{}^\frac{1}{2}/Z)}{d\alpha}\right)\;
\frac{\beta}{\lambda} +  m^2\,\frac{d\ln(Z_\lambda{}^\frac{1}{2}/Z)}{dm^2}\;\gamma_m &=&
\mu\,\frac{d\ln(Z/Z_\lambda{}^\frac{1}{2})}{d\mu}\cr & &
\eea
This is a inhomogeneous linear system with straightforward solution; at one loop
\bea
\beta|_{1l} &=&\lambda\;\mu\,\frac{d\ln( Z/Z_\lambda{}^\frac{1}{2})}{d\mu}\cr
\gamma_m|_{1l}&=& \mu\,\frac{d\ln(Z/Z_m)}{d\mu}\label{renfun1l}
\eea
Now from (\ref{actionpert}), (\ref{rencons}) we obtain counter-term contributions to the
two and three-point functions of the form
\bea
\tilde\Gamma_{a_1a_2}^{(1ct)}(p) &=& \left((Z-1)\;p^2 + (Z_m -1)\;
m^2\right)\;\kappa_{a_1a_2}\cr
\tilde\Gamma_{a_1a_2a_3}^{(1ct)}(\vp) &=& (Z_\alpha -1)\;\tilde\Gamma_{a_1a_2a_3}^{(0)}(\vp)
\eea
To renormalize the theory, from (\ref{unoa})-(\ref{g}) we fix at arbitrary $\mu$
\bea
Z|_{1l} &=& 1 + \frac{N\,\lambda^2}{24\,\pi}\;(1 - 3\,\alpha^2)\; \ln\frac{\Lambda^2}{\mu^2}\cr
Z_m|_{1l} &=& 1 + \frac{N\,\lambda^2}{24\,\pi}\;\left( 2\,\frac{\Lambda^2}{m^2} +
\ln\frac{\Lambda^2}{\mu^2}\right)
\eea
Similarly for the three-point function and taking into account the finiteness of contribution
$``(c)"$ we fix from (\ref{31a}), (\ref{31b})
\be
Z_\alpha|_{1l} = 1 + \frac{N\lambda^2}{16\,\pi}\;\ln\frac{\Lambda^2}{\mu^2} -
\frac{3\,N\lambda^2}{16\,\pi}\;\ln\frac{\Lambda^2}{\mu^2}
\ee
and admitting (\ref{Zalfa}) holds we read
\be
Z_\lambda|_{1l} = 1 + \frac{1}{2}\, (Z- 1) + \frac{1}{3}\, (Z_\alpha -1) =
1 - \frac{N\, \lambda^2}{24\,\pi}\; (1 + 3\,\alpha^2)\;\ln\frac{\Lambda^2}{\mu^2}
\ee
From (\ref{zlambda}), (\ref{renfun1l}) we finally get
\bea
\beta|_{1l} &=& -\frac{N\,\lambda^3}{8\,\pi}\;(1 - \alpha^2)\cr
\gamma_m|_{1l} &=& \frac{N\,\lambda^2\,\alpha^2}{4\,\pi}\cr
\gamma_\alpha|_{1l} &=& -\frac{N\,\lambda^2\,\alpha}{4\,\pi}\;(1-\alpha^2)\label{rgf}
\eea
They are essentially the same as in the commutative case but applied to the whole fields;
as we have pointed before in the commutative case they refer just to the $SU(N)$ part.
It is worth to note that the contributions to (\ref{rgf}) coming from the tadpole
diagrams from the two-point and three-point correlators exactly cancel leaving
the contributions from the graphs $(a)$ and $(b)$ respectively.
Furthermore if we introduce the renormalization constant $Z_g$ by \cite{boss}
\be
\frac{m_0{}^2}{\lambda_0{}^2} \equiv Z_g\;\frac{m^2}{\lambda^2} \;\;
\leftrightarrow\;\; Z_g = Z_\lambda{}^{-1}\; Z\; Z_m \stackrel{1l}{=} 1 +
\frac{N\, \lambda^2}{24\,\pi}\;( 2\,\frac{\Lambda^2}{m^2} +
3\,\ln\frac{\Lambda^2}{\mu^2})
\ee
then the anomalous dimension for the field $g$ is given by
\be
\gamma_g \equiv - \mu\frac{d\ln Z_g}{d\mu}\; \stackrel{1l}{=}\; \frac{N\,\lambda^2}{4\,\pi}\;
\stackrel{\alpha=1}{\longrightarrow}\; \frac{N}{k}
\ee
as we could hope from current algebra representation theory for a field transforming in the
$U(N)$ fundamental representation in the critical model \cite{goddol}.

\section{The one-loop four-points function}
\cleqn
In this section we will compute the coupling constant renormalization from the
four-points function verifying that (\ref{Zalfa}) indeed holds.

The tree level vertex is
\bea
\tilde\Gamma_{a_1\dots a_4}^{(0)}(\vp) &=& \frac{\lambda^2}{48}\; \kappa_{a_1\dots a_4}\;
e^{E_4(\vp)}\; (\Gamma(\vp) + 2\,m^2) + (perm.)\cr
E_4(\vp) &=& \exp\left( \frac{\theta}{2i}\,(\vp_1\times\vp_2 + \vp_3\times\vp_4)\right)\cr
\Gamma(\vp) &=& 2\,u^2 - s^2 - t^2
\eea
where we have introduced $\vs\equiv \vp_1 + \vp_2\;,\; \vu\equiv \vp_1 + \vp_3\;,\;
\vt\equiv \vp_1 + \vp_4\;, s^2+t^2+u^2=\sum_{i=1}^4 p_i{}^2$.

At one-loop there are six contributions to this correlator that we write below with their
divergent parts.

The contribution ``(a)" is a tadpole diagram with a six-points vertex,
\bea
\tilde\Gamma_{a_1\dots a_4}^{(1a)}(\vp)&=& \frac{\lambda^4}{5!\,3}\left(
-N\,\kappa_{a_1\dots a_4}\;\Gamma_1^{(1a)}(\vp) +
\kappa_{a_1}\,\kappa_{a_2a_3a_4}\;\Gamma_2^{(1a)}(\vp)+
\kappa_{a_1a_2}\,\kappa_{a_3a_4}\;\Gamma_3^{(1a)}(\vp)\right)\cr
&+& (perm.)\cr
\Gamma_i^{(1a)}(\vp) &=& \int\;\frac{d^2\vk}{4\,\pi^2}\; \tilde G (k^2)\;
\left\{\; \begin{array}{rl} v_c^6(-\vk,\vk,\vp_1,\dots,\vp_4)\;\;&,\;i=1\\
v_c^6(-\vk,\vp_1,\vk, \vp_2,\vp_3,\vp_4)\;\;&,\;i=2\\
\frac{1}{2}\; v_c^6(-\vk,\vp_1,\vp_2,\vk,\vp_3,\vp_4)\;\;&,\;i=3\end{array}
\right.
\eea
Its divergent part is given by
\bea
\tilde\Gamma_{a_1\dots a_4}^{(1a)}(\vp)|_{div.} &=& \frac{N\,\lambda^4}{48\cdot12\pi}\;
\kappa_{a_1\dots a_4}\; e^{E_4(\vp)}\;\left( -\frac{4}{5}\,\Lambda^2 + (\Gamma(\vp) +
2\,m^2)\;\ln\frac{m^2}{\Lambda^2}\right.\cr
&+& \left.(-\frac{6}{5}\,m^2 + \frac{1}{5}(s^2+t^2) - \frac{3}{10}\,u^2)\ln\frac{m^2}{\Lambda^2}
\right) + (perm.)
\eea

The contribution ``(b)" is a diagram with two four-points vertices,
\bea
\tilde\Gamma_{a_1\dots a_4}^{(1b)}(\vp)&=& -\frac{\lambda^4}{4!\,12}\left(
-N\,\kappa_{a_1\dots a_4}\;\Gamma_1^{(1b)}(\vp) +
\kappa_{a_1}\,\kappa_{a_2a_3a_4}\;\Gamma_2^{(1b)}(\vp)+
\kappa_{a_1a_2}\,\kappa_{a_3a_4}\;\Gamma_3^{(1b)}(\vp)\right.\cr
&+&\left. \kappa_{a_1a_4}\,\kappa_{a_2a_3}\;\Gamma_4^{(1b)}(\vp)\right) +(perm.)\cr
\Gamma_i^{(1b)}(\vp) &=& \int\;\frac{d^2\vk}{4\,\pi^2}\; \tilde G (k^2)\;\tilde G ((\vk-\vs)^2)
\left\{\; \begin{array}{r}
v_c^4(\vp_1,\vp_2, \vk,-\vk-\vs)\;v_c^4(\vp_3,\vp_4, -\vk+\vs,\vk)\\
2\; v_c^4(\vp_1,\vp_2, -\vk,\vk-\vs)\;v_c^4(\vp_3,\vk,\vp_4, -\vk+\vs)\\
v_c^4(\vp_1,\vp_2, -\vk,\vk-\vs)\;v_c^4(\vp_3,\vp_4, \vk, -\vk+\vs)\\
\frac{1}{2}\; v_c^4(\vp_1,-\vk,\vp_2,\vk-\vs)\;v_c^4(\vp_3,\vk,\vp_4, -\vk+\vs)\\
\end{array}\right.\cr
&&
\eea
for $i=1,2,3,4$ respectively.
Its divergent part is given by
\bea
\tilde\Gamma_{a_1\dots a_4}^{(1b)}(\vp)|_{div.} &=& \frac{N\,\lambda^4}{48\cdot12\pi}\;
\kappa_{a_1\dots a_4}\; e^{E_4(\vp)}\;\left( \frac{1}{2}\,\Lambda^2 -\frac{3}{2}\;
(\Gamma(\vp) + 2\,m^2)\;\ln\frac{m^2}{\Lambda^2}\right.\cr
&+& \left.(2\,m^2 + \frac{1}{2}\,u^2)\ln\frac{m^2}{\Lambda^2}
\right) + (perm.)
\eea

The contribution ``(c)" is a diagram with a three-points vertex and a four-points vertex,
\bea
\tilde\Gamma_{a_1\dots a_4}^{(1c)}(\vp)&=& \frac{\alpha^2\,\lambda^4}{180}\left(
-N\,\kappa_{a_1\dots a_4}\;\Gamma_1^{(1c)}(\vp) +
\kappa_{a_1a_2}\,\kappa_{a_3a_4}\;\Gamma_2^{(1c)}(\vp)+
\kappa_{a_4}\,\kappa_{a_1a_2a_3}\;\Gamma_3^{(1c)}(\vp)\right.\cr
&+&\left. \kappa_{a_3}\,\kappa_{a_1a_2a_4}\;\Gamma_4^{(1c)}(\vp)\right) +(perm.)\cr
\Gamma_i^{(1c)}(\vp) &=& \int\;\frac{d^2\vk}{4\,\pi^2}\; \tilde G (k^2)\;\tilde G ((\vk+\vp_4)^2)
\left\{\begin{array}{r}
v_c^3(\vk,\vp_4, -\vk-\vp_4)\;v_c^5(\vp_1,\vp_2,\vp_3, -\vk,\vk+\vp_4)\\
v_c^3(\vp_4,\vk, -\vk-\vp_4)\;v_c^5(\vp_1,\vp_2,-\vk,\vp_3, \vk+\vp_4)\\
v_c^3(\vp_4,\vk,-\vk-\vp_4)\;v_c^5(\vp_1,\vp_2,\vp_3, -\vk,\vk+\vp_4)\\
v_c^3(\vk,\vp_4,-\vk-\vp_4)\;v_c^5(\vp_1,\vp_2,-\vk,\vp_3, \vk+\vp_4)\\
\end{array}\right.\cr
&&
\eea
for $i=1,2,3,4$ respectively.
Its divergent part is given by
\be
\tilde\Gamma_{a_1\dots a_4}^{(1c)}(\vp)|_{div.} = \frac{N\,\alpha^2\lambda^4}{48\cdot 4\pi}\;
\kappa_{a_1\dots a_4}\; e^{E_4(\vp)}\;\left(-\Gamma(\vp) +
\vp_1\cdot\vp_3+\vp_2\cdot\vp_4\right)\;\ln\frac{m^2}{\Lambda^2} +
(perm.)
\ee

The contribution ``(d)" is a diagram with two three-points vertices and one four-points vertex,
\bea
\tilde\Gamma_{a_1\dots a_4}^{(1d)}(\vp)&=& -\frac{\alpha^2\,\lambda^4}{108}\left(
-N\,\kappa_{a_1\dots a_4}\;\Gamma_1^{(1d)}(\vp) +
\kappa_{a_1a_2}\,\kappa_{a_3a_4}\;\Gamma_2^{(1d)}(\vp)+
\kappa_{a_3}\,\kappa_{a_1a_2a_4}\;\Gamma_3^{(1d)}(\vp)\right.\cr
&+& \kappa_{a_4}\,\kappa_{a_1a_2a_3}\;\Gamma_4^{(1d)} +
\kappa_{a_2}\,\kappa_{a_1a_3a_4}\;\Gamma_5^{(1d)} +
\kappa_{a_1}\,\kappa_{a_2a_4a_3}\;\Gamma_6^{(1d)} +
\kappa_{a_1a_4}\,\kappa_{a_2a_3}\;\Gamma_7^{(1d)}\cr
&+&\left.\kappa_{a_1a_3}\,\kappa_{a_2a_4}\;\Gamma_8^{(1d)}\right) +(perm.)\cr
\Gamma_i^{(1d)}(\vp) &=& \int\;\frac{d^2\vk}{4\,\pi^2}\; \tilde G (k^2)\;
\tilde G ((\vk-\vs)^2)\;\tilde G ((\vk+\vp_3)^2)\;v_c^4(\vp_1,\vp_2, -\vk,\vk-\vs)\cr
& &
\left\{\; \begin{array}{rl}
v_c^3(\vk,\vp_3, -\vk-\vp_3)\;v_c^3(\vk+\vp_3,\vp_4, -\vk+\vs)\;\;\;\;,\;\;\;i=1\\
v_c^3(\vp_3,\vk, -\vk-\vp_3)\;v_c^3(\vp_4,\vk +\vp_3, -\vk+\vs)\;\;\;\;,\;\;\;i=2\\
v_c^3(\vp_3,\vk,-\vk-\vp_3)\;v_c^3(\vk+\vp_3,\vp_4, -\vk+\vs)\;\;\;\;,\;\;\;i=3\\
v_c^3(\vk,\vp_3,-\vk-\vp_3)\;v_c^3(\vp_4,\vk+\vp_3,-\vk+\vs)\;\;\;\;,\;\;\;i=4\\
\end{array}\right.\cr
&&
\eea
while for $i=5,6,7,8$ the expression is similar with $\;v_c^4(\vp_1,\vp_2, -\vk,\vk-\vs)\;$
replaced by $\;\frac{1}{2}v_c^4(\vp_1,-\vk,\vp_2,\vk-\vs)\;$.
Its divergent part is given by
\be
\tilde\Gamma_{a_1\dots a_4}^{(1d)}(\vp)|_{div.} = \frac{N\,\alpha^2\lambda^4}{48\cdot 4\pi}\;
\kappa_{a_1\dots a_4}\; e^{E_4(\vp)}\;\left(\frac{1}{2}\,\Gamma(\vp) -
\vp_1\cdot\vp_3-\vp_2\cdot\vp_4\right)\;\ln\frac{m^2}{\Lambda^2} +
(perm.)
\ee

The contribution ``(e)" is a diagram with four three-points vertices,
\bea
\tilde\Gamma_{a_1\dots a_4}^{(1e)}(\vp)&=& -\frac{\alpha^4\,\lambda^4}{81}\left(
\kappa_{a_1}\,\kappa_{a_2a_4a_3}\;\Gamma_1^{(1e)}(\vp)+
\kappa_{a_1a_4}\,\kappa_{a_2a_3}\;\Gamma_2^{(1e)}(\vp)+
\kappa_{a_1a_3}\,\kappa_{a_2a_4}\;\Gamma_3^{(1e)}(\vp)
\right.\cr
&-&\left.N\,\kappa_{a_1a_4a_3a_2}\;\Gamma_4^{(1e)}(\vp) +\right) +(perm.)\cr
\Gamma_i^{(1e)}(\vp) &=& \int\;\frac{d^2\vk}{4\,\pi^2}\; \tilde G (k^2)\;
\tilde G ((\vk-\vp_1)^2)\;\tilde G ((\vk+\vp_2)^2)\;\tilde G ((\vk-\vu)^2)\;
v_c^3(\vp_2,\vk,-\vk-\vp_2)\cr
& &\left\{\; \begin{array}{r}
v_c^3(\vp_1-\vk,\vp_4,\vk-\vt)\;v_c^3(\vp_3,\vk+\vp_2,-\vk+\vt)\;
v_c^3(-\vp_1,\vk,-\vk+\vp_1)\\
\frac{1}{2}\;v_c^3(\vp_4,\vp_1 -\vk,\vk-\vt)\;v_c^3(\vp_3,\vk+\vp_2, -\vk+\vt)\;
v_c^3(-\vp_1,\vk,-\vk+\vp_1)\\
\frac{1}{4}\;v_c^3(\vp_1 -\vk,\vp_4,\vk-\vt)\;v_c^3(\vk+\vp_2,\vp_3, -\vk+\vt)\;
v_c^3(-\vp_1,\vk,-\vk+\vp_1)\\
\frac{1}{4}\;v_c^3(\vp_1 -\vk,\vp_4,\vk-\vt)\;v_c^3(\vp_3,\vk+\vp_2, -\vk+\vt)\;
v_c^3(\vk,-\vp_1,-\vk+\vp_1)\\
\end{array}\right.\cr
&&
\eea
for $i=1,2,3,4$ respectively.
This contribution, as ``(c)" in the three-points function, is UV and IR finite for
general external momenta.

Finally from the definitions (\ref{rencons}) we get the counter-term contribution
\be
\tilde\Gamma_{a_1\dots a_4}^{(1ct)}(\vp) = (Z_\lambda -1)\;
\tilde\Gamma_{a_1\dots a_4}^{(0)}(\vp) +
\frac{Z_\lambda}{Z}\;(Z_m - Z)\; \frac{m^2\,\lambda^2}{4!}\;
\kappa_{a_1\dots a_4}\; e^{E_4(\vp)} + (perm.)
\ee

By summing up all the contributions we arrive to the result
\bea
\tilde\Gamma_{a_1\dots a_4}^{(1l)}(\vp)|_{div.} &=&
\frac{N\,\lambda^4}{48\cdot12\,\pi}\;\kappa_{a_1\dots a_4}\; e^{E_4(\vp)}\;
\left( \frac{17}{10}\,\Lambda^2 + \frac{1}{5}\;\sum_{i=1}^4 (p_i{}^2 + m^2)\right) +
(perm.)\cr
&+& \left(Z_\lambda|_{1l} - 1 - \frac{N\,\lambda^2}{24\,\pi}\; (1 + 3\,\alpha^2)\;
\ln\frac{m^2}{\Lambda^2}\right)\;\tilde\Gamma_{a_1\dots a_4}^{(0)}(\vp)\label{Zlambda}
\eea
The second line line fixes the one-loop value of $Z_\lambda$ exactly to that given in
(\ref{Zalfa}) verifying in this way the non-renormalization of the level $k$,
while the second term in the first line needs a non left-right invariant counter-term of
the form
\be
\delta S[\pi] =\frac{N\,\lambda^4}{48\cdot 15\,\pi}\;\ln\frac{\Lambda^2}{\mu^2}\;
\int d^2\vx\;Tr\,\left( \pi^3\;(-\Box +m^2)\pi\right)\label{ctoffs}
\ee
which however is identically zero on-shell!

\section{Conclusions}

We have computed in a Schwinger regularization scheme (not usual to my knowledge in the
literature on the subject) correlation functions in a non commutative version of the
two dimensional WZW model, where other than the mass parameter another dimensional one,
the $\theta$ parameter, is present.
The results display some features common to other field theories with non-derivative
vertices, in particular the different behavior of them depending on the range of the
parameters as made explicit in Appendix B and (\ref{g}), which is at the origin of the
so-called UV/IR mixing.
As a by product we computed the renormalization group functions in a setting less involved
than the dimensional regularization context; to this respect we would
like to spend some words about the background field method.
In order to carry out computations in this context from (\ref{efac}) and taking the standard
splitting other than the vertices seen new ones appear from (\ref{polwieg})
\be
\frac{1}{\lambda^2}\;\int d^2\vx\;
Tr\left(\,P_{-}^{ij}\; L_i(g_0) \; R_j(e^{\lambda\,\pi})\right)\label{verefac}
\ee
The relevant term at one loop coming from (\ref{verefac}) is
\be
V_2 =\frac{1}{2}\;\int d^2\vx\;
Tr\left(\,P_{-}^{ij}\; L_i(g_0) \; [\partial_j\,\pi,\pi]_*\right)\
\ee
Then the quadratic contribution to the effective action is given by
\bea
\Gamma_2^{(1)}[g_0] &=& -\frac{1}{2}\;<V_2\;V_2>_0\cr
&=&\frac{1}{8}\,P_{-}^{ij}\,P_{-}^{kl}\;\int d^2\vx\int d^2\vy\;
L_i^{a_1}(g_0)|_{\vx}\; L_k^{a_2}(g_0)|_{\vx-\vy} \;
\left(-N\,\kappa_{a_1a_2}\; f_{jl}(y;0)\right.\cr
&-&\left. \kappa_{a_1}\,\kappa_{a_2}\; f_{jl}(y;\theta)\right)\cr
f_{jl}(y;\theta) &=& \int\,\frac{d^2\vk_1}{4\pi^2}\,\int\,\frac{d^2\vk_2}{4\pi^2}\;
\tilde G(k_1)\;\tilde G(k_2)\; (\vk_1-\vk_2)_j\;(\vk_1-\vk_2)_l\;
e^{i(\vk_1+\vk_2)\cdot\vy -i\theta\vk_1\times\vk_2}\label{efacdos}
\eea
This is essentially the computation carried out recently in reference $\cite{fur}$
(and time ago in \cite{wit} in the commutative case); we do not reproduce the details here,
just to say that as verification if we carry out (\ref{efacdos}) and renormalize we get
exactly the beta function giving in (\ref{rgf}).
\footnote{
We should say that we do not agree in the denominator with the result of reference
\cite{fur} that is $32\pi$; we get $8\pi$ instead.
}
The fact that we obtain the same non-trivial fixed point at the critical point $\alpha=1$
as well as the right conformal dimension in the fundamental representation for the field $g$
seems to indicate that both theories in the critical point could be equivalent.
This fact is supported also for the non renormalization of $k$ here verified explicitly
and by recent computations of fermionic determinants \cite{fiduno} and a kind of
Seiberg-Witten map recently proposed \cite{fiddos} that would prove the equivalence.
If this is so then we would certainly have examples of {\it unitary} field theories NC in
time directions, a very different behaviour of non derivative scalar theories recently
studied in \cite{gomis3}.

In what the four-points correlator computed in Section $6$ concerns we would like to comment
a couple of things.
The first one is the presence of quadratic divergences (c.f equation (\ref{Zlambda})),
a natural fact in our regularization context.
They should be killed by terms coming from a left-right invariant measure \cite{peter},
\cite{leut}; this is a very subtle subject in the commutative case and it is more in
the non-commutative context; we just say that we have carried out all the computations made
above in the dimensional regularization scheme of reference \cite{boss} where the measure
problems are not present \cite{brezinn} because of the identity
\be
\int d^d\vk = 0  \;\;\;, \;\; d\equiv 2 + 2\epsilon
\ee
and we have obtained exactly the same results presented here with the
replacement of $\ln\Lambda^2$ with $1/\epsilon$ and the absence of quadratic divergences.

The second one is the need of (\ref{ctoffs}) to renormalize off-shell the theory.
While is worth to note that (no) tadpole contributions to $Z_\lambda$ in (\ref{Zlambda})
correspond to (no) tadpole contributions in (\ref{Zalfa}), both of them conspire in a non
trivial way to give this term.
We remark that it must be present in the commutative case also, and what is more, at the
level of the non-linear sigma-model obtained by putting $\alpha\equiv0$, because it comes
from the only two contributions (the diagrams ``(a)" and ``(b)") to the four-point correlator
with no odd vertices.
Its presence has the same origin as the presence of non-covariant terms in the effective action
of general two-dimensional sigma-models noted in references \cite{lagfm}, \cite{hull}, when the
expansion is made using a non-covariant field; in fact it is easy to show that in the present
context we can absorbe it by the field redefinition
\be
\pi(\vx) = \xi(\vx) + \frac{N\,\lambda^4}{48\cdot30\,\pi}\;\xi(\vx)^3 + o(\xi^5)
\ee

\section*{Acknowledgements}

We would like to thank Daniel Cabra, Enrique Moreno  and Fidel Schaposnik for
enlightening discussions and references pointed out.

\appendix

\section{Conventions}

We resume here the conventions adopted and some useful formulae.

The Moyal product is defined by
\be
f*g\,(\vx) \equiv \exp\left(\frac{i}{2}\, \theta^{\mu\nu} \frac{\partial}{\partial y^\mu}
\frac{\partial}{\partial z^\mu}
\right)\; f(\vy)\; g(\vz)\; \left|_{\vy=\vz=\vx}\right.\label{moyal}
\ee
The space of definition is $\Re^d$ because only on this manifold definition (\ref{moyal})
has a covariant meaning in cartesian coordinates.
\footnote{
Compactifications are certainly possible, e.g. the NC torus and few other examples \cite{cds}.
}
The functions are taken to be of integrable square; then in momentum space we have
\be
f_1 *\dots *f_m\, (\vx) = \int \frac{d^d \vk_1}{(2\pi)^d}\dots \frac{d^d \vk_m}{(2\pi)^d}\;
{\tilde f}_1(\vk_1)\dots {\tilde f}_m(\vk_m)\;
e^{i\, \vx\cdot\sum_{l=1}^m\, \vk_l - \frac{i}{2}\, \sum_{l<s=1}^m
\theta^{\mu\nu}\,k_{l\mu}\,k_{s\nu}}
\ee
It is evident that the reality of $\theta^{\mu\nu}$ is a necessary condition for the
existence of the Moyal product.
On the other hand the anti-symmetry of it allows to write
\bea
[f,g]_*(\vx) &\equiv& f*g\, (\vx) - g*f\, (\vx) = i\,\theta^{\mu\nu}\;\partial_\mu
J_\nu^-(\vx)\cr
\{f,g\}_*(\vx) &\equiv& f*g\, (\vx) + g*f\, (\vx) = 2\, f(\vx)\,g(\vx) + \theta^{\mu\nu}\;
\partial_\mu J_\nu^+(\vx)\label{moyalbra}
\eea
where the currents are given by
\bea
J_\mu^- (\vx) &=& \sum_{m\geq 0}\; \frac{(-)^m}{2^{2m+1}\, (2m+1)!}\,
\theta^{\mu_1\nu_1}\dots \theta^{\mu_{2m}\nu_{2m}}\;
\partial_{\mu_1}\dots\partial_{\mu_{2m}}\, f(\vx)\; \partial_\mu\,
\partial_{\nu_1}\dots\partial_{\nu_{2m}}\, g(\vx)\cr
&-& (f\leftrightarrow g)\cr
J_\mu^+ (\vx) &=& \sum_{m\geq 1}\; \frac{(-)^m}{2^{2m}\, (2m)!}\,
\theta^{\mu_1\nu_1}\dots \theta^{\mu_{2m-1}\nu_{2m-1}}\;
\partial_{\mu_1}\dots\partial_{\mu_{2m-1}}\, f(\vx)\; \partial_\mu\,
\partial_{\nu_1}\dots\partial_{\nu_{2m-1}}\, g(\vx)\cr
&+& (f\leftrightarrow g)\label{moyalcur}
\eea
The fact that the Moyal bracket is a total derivative is a fundamental property  allowing
integration by parts and cyclic properties under integration to hold.

Let $\pi = \pi^a\, X_a \in \cal G$ where $\{X_a\}$ are some Lie algebra generators; then
\be
[\pi_1 ,\pi_2 ]_* = \frac{1}{2}\, \{ \pi_1^a , \pi_2^b\}_*\; [X_a , X_b] +
\frac{1}{2i}\, [ \pi_1^a , \pi_2^b]_*\; i\,\{X_a , X_b\}
\ee
that from properties (\ref{moyalbra}), (\ref{moyalcur}) closes in the algebra only for
${\cal G} = gl(N,C)$ or its restriction $u(N)$.
The generators of $u(N)$ can be taken as those of $su(N)$ plus the identity, and in the
paper is used the definition
\bea
\kappa_{a_1\dots a_n} &\equiv& Tr\,\left( X_{a_1}\dots X_{a_n}\right)\cr
Tr\,\left(\, \dots \right) &\equiv& - tr_{F}\,\left( \dots\right)
\eea
where the last line defines the scalar product adopted (denoted by ``Tr") and ``F" stands
for the fundamental $N$-dimensional representation.
In particular $\kappa_{ab}$ is the metric used to rise and low indices in the algebra.
However it is more convenient to expand in terms of the generators of $gl(N,C)$, the matrices
$(E_{ij})_{kl}= \delta_{ik}\,  \delta_{jl}$.
In this basis we must have in mind that $\pi^{ij\,*} = - \pi^{ji}$.
By using this basis we easily get the various useful formulae
\bea
\kappa_{a_1}{}^{bc}\;\kappa_{a_2cb} &=&
\kappa_{a_1a_2b}{}^b =-N\,\kappa_{a_1 a_2} \cr
\kappa_{a_1}{}^{bc}\;\kappa_{a_2bc} &=&
\kappa_{a_1ba_2}{}^b =\kappa_{a_1}\,\kappa_{a_2}\cr
\kappa_{a_1a_2a_3b}{}^b &=&\kappa_{a_1}{}^{bc}\,\kappa_{a_2a_3cb}=
\kappa_{a_1}{}^{bc} \kappa_{a_2db}\,\kappa_{a_3c}{}^d= -N\,\kappa_{a_1a_2a_3}\cr
\kappa_{a_1a_2ba_3}{}^{b} &=&\kappa_{a_1}{}^{bc}\,\kappa_{a_2ca_3b} =
\kappa_{a_1}{}^{bc} \,\kappa_{a_2db}\,\kappa_{a_3}{}^d{}_c=
\kappa_{a_1a_2}\;\kappa_{a_3}\cr
\kappa^b{}_{ba_1\dots a_4}&=&
\kappa_{a_1a_2}{}^{cb}\;\kappa_{a_3a_4bc}=
\kappa_{a_4bc}\;\kappa_{a_1a_2a_3}{}^{cb}=
\kappa_{a_1a_2b}{}^c\;\kappa_{a_3}{}^{db}\;\kappa_{a_4cd}\cr
&=&\kappa_{a_1}{}^{be}\;\kappa_{a_2db}\;\;\kappa_{a_3}{}^{cd}\;\kappa_{a_4ec}=
-N\,\kappa_{a_1\dots a_4}\cr
\kappa^b{}_{a_1ba_2a_3a_4}&=&
\kappa_{a_1}{}^b{}_{a_2}{}^{c}\;\kappa_{a_3a_4bc}=
\kappa_{a_1bc}\;\kappa_{a_2a_3a_4}{}^{bc}=
\kappa_{a_4bc}\;\kappa_{a_2a_3}{}^c{}_{a_1}{}^b=
\kappa_{a_2a_3b}{}^c\;\kappa_{a_1}{}^{bd}\;\kappa_{a_4cd}\cr
&=&
\kappa_{a_2a_3b}{}^c\;\kappa_{a_4}{}^{db}\;\kappa_{a_1dc}=
\kappa_{a_1ba_2c}\;\kappa_{a_4}{}^{bd}\;\kappa_{a_3dc}=
\kappa_{a_4ba_1c}\;\kappa_{a_2}{}^{db}\;\kappa_{a_3cd}\cr
&=&\kappa_{a_1}{}^{be}\;\kappa_{a_2bc}\;\kappa_{a_4}{}^{cd}\;\kappa_{a_3de}=
\kappa_{a_1}\;\kappa_{a_2a_3a_4}\cr
\kappa^b{}_{a_1a_2ba_3a_4}&=&
\kappa_{a_1a_2}{}^{bc}\;\kappa_{a_3a_4bc}=
\kappa_{a_1}{}^b{}_{a_4}{}^{c}\;\kappa_{a_3ba_2c}=
\kappa_{a_4}{}^{bc}\;\kappa_{a_1a_2ba_3c}=
\kappa_{a_1a_2b}{}^c\;\kappa_{a_3}{}^{bd}\;\kappa_{a_4dc}\cr
&=&
\kappa_{a_1ba_4c}\;\kappa_{a_3}{}^{bd}\;\kappa_{a_2cd}=
\kappa_{a_1ba_3c}\;\kappa_{a_2}{}^{db}\;\kappa_{a_4dc}=
\kappa_{a_1}{}^{be}\;\kappa_{a_4bc}\;\kappa_{a_3}{}^{cd}\;\kappa_{a_2ed}\cr
&=&\kappa_{a_1}{}^{be}\;\kappa_{a_3bc}\;\kappa_{a_2}{}^{dc}\;\kappa_{a_4de}=
\kappa_{a_1a_2}\;\kappa_{a_3a_4}
\eea

In terms of $g\equiv \exp(\lambda\pi)$ the NC left and right invariant Maurer-Cartan
$u(N)$-valued forms have the expansions
\bea
L(g) &\equiv& g^{-1}\; dg = \sum_{m\geq 1} \, \frac{\lambda^m}{m!}\; \omega_m(\pi)
\;\;\;\;\;\;\;\;\;\;\;\;,\;\;\; dL(g) + L(g)\wedge L(g) = 0\cr
R(g) &\equiv& dg\; g^{-1} = -\sum_{m\geq 1} \, \frac{(-\lambda)^{m}}{m!}\;\omega_m(\pi)
\;\;\; , \;\;\; dR(g) - R(g)\wedge R(g) = 0\cr
\omega_m(\pi) &=& \sum_{p=1}^m\,(-)^{p+1}\;\left( \begin{array}{c}m\\p\end{array}\right)\;
\sum_{k=1}^p\; \pi^{k-1}\; d\pi\; \pi^{m-k} =
[\dots[[d\pi ,\overbrace{\pi]_*,\pi]_*,\dots ,\pi ]_*}^{m-1}\cr
& &
\eea
The even vertices of the action (\ref{accion}) come from $I_0$
while the odd ones come from the WZ term; however it is better to work out them in a
unified way following \cite{boss}.
To this end we consider
\bea
S[e^{\lambda\,\pi}] &=& \int_0^\lambda\;dt\;\frac{dS[e^{t\,\pi}]}{dt}=
\int_0^\lambda\;dt\; \lim_{\Delta t \rightarrow 0}\;\frac{1}{\Delta t}\;
(S[g + \delta g] - S[g])|_{\delta g = \Delta t\, g\,\pi\,,\; g=e^{t\pi}}\cr
&=& \frac{1}{\lambda^2}\; \int^\lambda_0\; dt\;\int d^2\vx\; P_+^{ij}\;
Tr\left( \partial_i\pi\, L_j(e^{t\,\pi})\right)
\eea
where in the last line we used (\ref{varac}).
From here we read the vertices in the form
\be
S^n[\pi] = \frac{\lambda^{n-2}}{n!}\;\int d^2\vx\; P_+^{ij}\;
Tr\left( \partial_i\pi\; \omega_{n-1,j}(\pi)\right)\;\;\; ,\;\; n\geq 3
\ee

\section{Computation of $f(p;\theta)$}

In this appendix we describe the exact computation of a one-loop integral; others integrals
in the paper, more or less involved they be, are computed following similar same steps.
We would like to point out that in the context of dimensional regularization the introduction of
Schwinger-Feynman parametrizations is the key to evaluate integrals; however in our
regularization procedure they are not so useful due to the presence of the exponential factors
in the propagators, so we must follow another route.

The function introduced in Section $3$ is given by
\footnote{
Dependence in the cut - off parameters are systematically omitted.
}
\be
f(p;\theta) = \int\, \frac{d^2\vk}{(2\pi)^2}\;
\frac{\exp\left( -\frac{1}{2\Lambda^2}\,\left( \vk^2 + (\vk-\vp)^2\right) + i\,\theta\,
\vp\times\vk\right)}{(\vk^2+m^2)\;((\vk-\vp)^2 +m^2)}\;(\vp\times\vk)^2\label{fcero}
\ee
It is clear almost by definition that is null at $p=0$; then in the computation and by using
rotational invariance we will take $\vp = p\,\check n\,$ with $p>0\,,\check n=(1,0)$, and the
final result will take into account the mentioned fact.
It will be convenient in what follows to introduce the dimensionless parameters
(do not confound $\mu$ with any free scale)
\be
\begin{array}{rclcl}
\sigma_\pm &\equiv& \sigma\pm 1  &,&\;\;\;\sigma\equiv\theta\,\Lambda^2\cr
x_0 &\equiv& \frac{1}{2}\, (\mu^2 +
\sigma_+\,\sigma_-)&,&\;\;\;\mu\equiv\frac{2m}{p}\cr
x_\pm &\equiv& \sigma^2 \pm\sqrt{\sigma^2- \mu^2}&,&\;
x_m \equiv \mu\,\sqrt{\sigma_+\,\sigma_-}
\end{array}
\ee
By making the change of variables $\vx = \frac{2}{p}\,\vk - \check n$ we write (\ref{fcero}) as
\be
f(p;\theta) = e^{-\frac{p^2}{4\Lambda^2}\left( 1+\sigma^2\right) }\;
\frac{p^2}{2\pi}\;\int\,\frac{d^2\vx}{2\pi}\;
\frac{\exp\left( -\frac{p^2}{4\Lambda^2}\,(\vx + i\,\sigma \epsilon\check n)^2\right)
\;(\check n\times\vx)^2}{\left( (\vx +\check n)^2 + \mu^2\right)\;
\left( (\vx-\check n)^2 + \mu^2\right)}\label{funo}
\ee
Now we must be careful because if the shift in a real vector just made is allowed, it is
not so in general (and here in particular).
In fact the integrand has poles in the complex $x^2$-plane and we must take them into account.
Explicitly we can write
\be
f(p;\theta) = -
\frac{ e^{\frac{m^2}{\Lambda^2} -\frac{p^2}{2\Lambda^2}}}{8\pi\,\sigma_+\,\sigma_-}\;
p^2\left( f_s(p;\theta) + f_r(p;\theta)\right)\label{f}
\ee
where $f_s(p;\theta)$ is the shifted integral
\be
f_s(p;\theta) = \sigma_+\,\sigma_-\;
\int\,\frac{d^2\vx}{2\pi}\;
\frac{ e^{-\frac{p^2}{4\Lambda^2}\left(\vx^2 + 2\,x_0\right)}
\;(2\,i\,\check n\times\vx -2\,\sigma)^2}{\left( (\vx  -i\,\sigma
\epsilon\check n+\check n )^2 + \mu^2\right)\;
\left( (\vx-i\,\sigma \epsilon\check n-\check n)^2+ \mu^2\right)}
\ee
and the contribution from the residues of the poles included in the strip\\
$S=\{w=w_1+iw_2\;: \; w_1\in\Re, w_2\in [0,\sigma]\}$ is
\bea
f_r(p;\theta) &=&  \int_{-\infty}^\infty\; dy\;\sum_{w_j\in S}\, Res(F(w;y); w_j)\cr
F(w;y) &=& -4\,\sigma_+\,\sigma_-\;
\frac{ i\, w^2\; \exp\left(-\frac{p^2}{4\Lambda^2}\,\left( w^2 -
2\,i\,\sigma\; w + y^2 + \mu^2 -1 \right)\right)}{\left(w^2 + (y+1)^2 +\mu^2\right)\;
\left(w^2 + (y-1)^2+\mu^2\right)}
\eea
In what $f_s$ concerns we introduce polar coordinates for $\vx = (x^1 , x^2)$ in the way
\be
x^1 + i\, x^2 = z\; \sqrt{2\, (x - x_0)} \;\;\; , \;\;
x\in [x_0 ,\infty)\;\;\;,\;\;|z|^2=1
\ee
Then after a re-scaling in $z$ we write
\bea
f_s (p;\theta) &=& \int_{x_0}^\infty\,dx\;
e^{-\frac{p^2}{2\Lambda^2}\, x}\; I_s(x)\cr
I_s(x) &=& \oint_{|z|^2=2(x-x_0)}\,\frac{dz}{2\pi i}\;
\frac{\left(z^2 - 2\,\sigma\, z- 2\,x + 2\,x_0\right)^2}
{z\,(z- z_{++}(x))\,(z-z_{-+}(x))\,(z-z_{+-}(x))\,(z-z_{--}(x))}\cr
& &
\eea
The contour integral $I_s(x)$ is then computed by using Cauchy theorem by evaluating
the residues of the integrand in $\,\{ z_0 = 0,\;z_{\pm+}(x)= -\frac{a_\mp (x)}{\sigma_+},\;
z_{\pm-}(x)= -\frac{a_\pm (x)}{\sigma_-}\}\;\;$ where
\be
a_{\pm}(x) = x - \sigma_+\,\sigma_-\pm\sqrt{ x^2-x_m{}^2}
\ee
All these poles lie on the real line and their residues are
\bea
Res(z_0) &=& 1\cr
Res(z_{\pm+}(x)) &=& -Res(z_{\pm-}(x))
=\mp \, \frac{1}{2}\;\frac
{\left(\sigma_+\, a_\pm(x) + \sigma_-\,a_\mp(x) + 2\,\sigma\,\sigma_+\,\sigma_- \right)^2}
{\left(a_+(x)-a_-(x)\right)\left(\sigma_+\, a_\pm(x) - \sigma_-\,a_\mp(x)  \right)}\cr
&=& \mp\;\frac{1}{2}\;\frac{(\sigma\, x \pm \sqrt{x^2 - x_m{}^2})^2}{\sqrt{x^2 - x_m{}^2}\;
\left(x - \sigma_+\,\sigma_- \pm\sigma\,\sqrt{x^2 - x_m{}^2}\right)}
\eea
Of particular relevance will be the sum
\bea
R(x) &\equiv& Res(z_{++}(x)) + Res(z_{--}(x))\cr
&=&  \left( x^2 -\mu^2\,\sigma_+\,\sigma_-\right)^{-\frac{1}{2}}\;
\left(-x - \sigma_+\,\sigma_- +\frac{\sigma_+\,\sigma_-}{2}\;\frac{\mu^2+r_+}{x+r_+}
+\frac{\sigma_+\,\sigma_-}{2}\;\frac{\mu^2+r_-}{x+r_-}\right)\cr
r_\pm &\equiv& 1\pm\sigma\;\sqrt{1+\mu^2}\label{R}
\eea
The next step is to analyze the contributions of theses residues.
From it we get
\be
f_s(p;\theta) =\left\{ \begin{array}{lcr}
\int_{x_0}^\infty\,dx\;e^{-\frac{p^2}{2\Lambda^2}\, x}\;
\left(1 + Res(z_{++}(x)) + Res(z_{--}(x))\right) &,&\;\;\sigma^2 - \mu^2<0\cr
\int_{x_0}^{x_-} \,dx\;e^{-\frac{p^2}{2\Lambda^2}\, x}\;
\left(1 + Res(z_{++}(x)) + Res(z_{--}(x))\right)\cr
+ \int_{x_-}^{x_+}\,dx\;e^{-\frac{p^2}{2\Lambda^2}\, x}\;
\left(1 + Res(z_{++}(x)) + Res(z_{-+}(x))\right)\cr
+ \int_{x_+}^\infty\,dx\;e^{-\frac{p^2}{2\Lambda^2}\, x}\;
\left(1 + Res(z_{++}(x)) + Res(z_{--}(x))\right)&,&\;\; 0<\sigma^2 - \mu^2<1\cr
\int_{x_0}^{x_-} \,dx\;e^{-\frac{p^2}{2\Lambda^2}\, x}\;
\left(1 + Res(z_{+-}(x)) + Res(z_{-+}(x))\right)\cr
+ \int_{x_-}^{x_+}\,dx\;e^{-\frac{p^2}{2\Lambda^2}\, x}\;
\left(1 + Res(z_{++}(x)) + Res(z_{-+}(x))\right)\cr
+ \int_{x_+}^\infty\,dx\;e^{-\frac{p^2}{2\Lambda^2}\, x}\;
\left(1 + Res(z_{++}(x)) + Res(z_{--}(x))\right)&,&\;\;\sigma^2 - \mu^2>1\cr
&\label{fs}
\end{array}\right.
\ee
In what $f_r$ concerns, the poles of $F(w;y)$ are localized in
$w_{\pm+}=\pm i\,\sqrt{(y+1)^2 + \mu^2}$ and $w_{\pm-}=\pm i\,\sqrt{(y-1)^2 + \mu^2}$.
Clearly only $w_{++}$ and $w_{+-}$ can lie inside the strip and they will contribute iff
$|w_{+\pm}|^2<\sigma^2$.
The result of this analysis yields
\be
f_r(p;\theta) = \left\{\begin{array}{lcr}
0&,& \sigma^2 -\mu^2<0\\
\sigma_+\,\sigma_-\;\int_{-\sqrt{\sigma^2-\mu^2}}^{\sqrt{\sigma^2-\mu^2}}\; dy\;
\frac{\sqrt{y^2 + \mu^2}}{y+1}\; e^{-\frac{p^2}{2\Lambda^2}\left(y + \sigma\,
\sqrt{y^2 + \mu^2}\right)} &,&\sigma^2-\mu^2>0
\end{array}\right.
\ee
We then make in the last case the change of variable
\be
x = y + \sigma\; \sqrt{ y^2 + \mu^2}
\ee
being careful with the inverse $y(x)$ to be considered that depends on the range of the
parameters.
We get
\be
f_r(p;\theta) =
\left\{\begin{array}{lcr}0&,&\sigma^2-\mu^2<0\\
\int_{x_-}^{x_+}\,dx\;e^{-\frac{p^2}{2\Lambda^2}\, x}\;
(-2\, Res(z_{-+}(x))&,&0<\sigma^2 - \mu^2<1\\
\int_{x_m}^{x_-}\,dx\;e^{-\frac{p^2}{2\Lambda^2}\, x}\;
\left(2\, Res(z_{++}(x) - 2\, Res(z_{-+}(x)\right)\\
+\;\;\int_{x_-}^{x_+}\,dx\;e^{-\frac{p^2}{2\Lambda^2}\, x}\;
\left(-2\, Res(z_{-+}(x)\right)&,&\sigma^2 - \mu^2>1
\end{array}\right.\label{fr}
\ee
From (\ref{f}), (\ref{fs}) and (\ref{fr}) we finally obtain
\be
f(p;\theta)=-\frac{ e^{\frac{m^2}{\Lambda^2}-
\frac{p^2}{2\Lambda^2}}}{8\pi\,\sigma_+\,\sigma_-}\;p^2\;\times
\left\{\begin{array}{lr}
\int_{x_0}^{\infty}\,dx\;e^{-\frac{p^2}{2\Lambda^2}\, x}\;
(1 + R(x))&,\;\; \sigma_+\,\sigma_- <\mu^2\\
\int_{x_m}^{x_0}\,dx\;e^{-\frac{p^2}{2\Lambda^2}\, x}\;(-1 + R(x))&\\
+ \int_{x_m}^{\infty}\,dx\;e^{-\frac{p^2}{2\Lambda^2}\, x}\;
(1 + R(x))&,\;\; \sigma_+\,\sigma_- >\mu^2
\end{array}\right.\label{ffinal}
\ee
The reader should note the different form it displays in the two regions of parameters,
a kind of phase transition around the value
$\theta_c = \frac{\sqrt{p^2 + 4\,m^2}}{\Lambda^2\,p}$.
In particular if we like its value for $\theta=0$ we must use the first expression;
in the massless limit we get
\bea
\lim_{m\rightarrow 0}\,f(p;0) &=& \frac{p^2}{8\pi}\; \left( -2\,
Ei\left(-\frac{p^2}{2\Lambda^2}\right) +
Ei\left(-\frac{p^2}{4\Lambda^2}\right) + e^{-\frac{3}{8}\frac{p^2}{\Lambda^2}}\;
\frac{\sinh(\frac{p^2}{8\Lambda^2})}{\frac{p^2}{8\Lambda^2}}\right)\cr
& &\;\stackrel{\Lambda\rightarrow\infty}{\longrightarrow}\;\frac{p^2}{8\pi}\;
\left( -\ln\frac{p^2}{\Lambda^2} +1-\gamma \right)
\eea
where $Ei(x)$ is the exponential-integral function and $\gamma$ the Euler constant
(see \cite{grad}).
Note that for $p=0$ is null as should from the remark at the beginning;
furthermore it displays a logarithmic singularity.

On the other hand, if we like its value at $\theta\neq 0$ for large enough cut-off
$\Lambda$ we must consider the second expression; in the massless limit we get
\bea
\lim_{m\rightarrow 0}\,f(p;\theta) &=& \frac{p^2}{8\,\pi}\; \left(
e^{-\frac{\sigma^2 + 3}{8}\,\frac{p^2}{\Lambda^2}}\;
\frac{\sinh\frac{\sigma_+\sigma_-p^2}{8\Lambda^2}}{\frac{\sigma_+\sigma_-p^2}{8\Lambda^2}}
+ e^{\frac{\theta\,p^2}{2}}\;
\left( - Ei( -\frac{\sigma_+ p^2}{2\Lambda^2})
+ \frac{1}{2}\,Ei( -\frac{\sigma_+{}^2 p^2}{4\Lambda^2}\right)\right.\cr
&+& \left. e^{-\frac{\theta\,p^2}{2}}\;\left( - Ei( -\frac{\sigma_- p^2}{2\Lambda^2})
+ \frac{1}{2}\,Ei( -\frac{\sigma_-{}^2 p^2}{4\Lambda^2}\right)
\right)\cr
& &\stackrel{\Lambda\rightarrow\infty}{\longrightarrow}
-\frac{p^2}{8\,\pi}\; \left(
e^{\frac{\theta\,p^2}{2}}\; Ei(-\frac{\theta\,p^2}{2})+
e^{-\frac{\theta\,p^2}{2}}\; Ei(\frac{\theta\,p^2}{2})
\right)
\eea
getting a finite result instead.
Some remarks are in order.
The first one concerns the massless limit; it is possible to prove that the results just obtained
coincide with the ones we had obtained by putting $m=0$ since the beginning.
And with respect to the limit $\Lambda\rightarrow\infty$, the computation at
$\theta\neq 0$ without cut-off agrees with it also.
The second one is connected  with the trick used to carry out the computations.
There is a particular case, maybe of little physical interest, in which the computations
simplifies a lot and corresponds to the scaling of the non commutative parameter defined by
$\theta\,\Lambda^2 =1$.
In this case (for simplicity we consider the massless limit) $f$ is written as
\be
f(p;\theta) = e^{-\frac{\theta p^2}{2}}\; \frac{p^2}{2\pi}\; \int_0^\infty\,
dk\,k\;e^{-\theta\, p^2\,k^2}\;\int_{-\pi}^\pi\, \frac{d\phi}{2\pi}\;
\frac{e^{\theta\, p^2\,k\,  e^{i\phi}}}{k^2 + 1 - 2\, k\, \cos\phi}\; \sin^2\phi
\label{fsigmauno}
\ee
Going to the complex variable  $z=e^{i\phi}$ the integral is straightforward and we finally
get
\be
f(p;\theta) = \frac{p^2}{8\pi}\; e^{-\frac{\theta p^2}{2}}\;
\left(1 - \frac{\gamma}{2} - \ln\sqrt{\theta\, p^2} -
\frac{e^{\theta p^2}}{2}\;Ei\left(-\theta p^2\right)\right)
\ee
This result can be obtained from (\ref{ffinal}) as a special case;
the important thing to observe is that the simplification in (\ref{fsigmauno}) occurs
because we remain with just $z$ in the exponent, in this way we remain with a meromorphic
function (outside $\infty$) and we can apply Cauchy theorem.
In the general case we have $z$ and $\frac{1}{z}$, and this last factor represents an essential
singularity at the origin which does not allow the residues calculation.
This is the reason because it is necessary to make the shift in the exponential which
leads to (\ref{f}) in order to eliminate any $z$ dependence.
We finally remark that this approach works in the presence of any rational function of
$\sin\phi$ and $\cos\phi$, in particular to compute any one-loop graph.



\begin{thebibliography}{99}

\bibitem{seiwit} N. Seiberg and E. Witten, JHEP {\bf 09} (1999), 32.

\bibitem{cds} A. Connes, M. Douglas and A. Schwarz: ``Non commutative geometry and
matrix theory: compactification on tori", hep-th/9711162.

\bibitem{luis} L. Alvarez-Gaum\'{e}  and S. Wadia: ``Gauge theory on a quantum phase space",
hep-th/0006219.

\bibitem{minraamsei} S. Minwalla, M. Van Raamsdonk and N. Seiberg:
``Non commutative perturbative dynamics", hep-th/9912072;
M. Van Raamsdonk and N. Seiberg: ``Comments on non commutative perturbative dynamics",
hep-th/0002186.

\bibitem{gomis2} J. Gomis, K. Landsteiner and E. Lopez: ``Non-relativistic
non-commutative field theory and UV/IR mixing", hep-th/0004115.

\bibitem{wit} E. Witten, Comm. Math. Phys. {\bf 92} (1984), 455.

\bibitem{dabro} L. Dabrowski, T. Krajewski and G. Landi: ``Some properties of non-linear
$\sigma$ models in noncommutative geometry", hep-th/0003099.

\bibitem{gomis1} J. Gomis, K. Kamimura and J. Llosa: ``Hamiltonian formalism for
space-time non commutative theories", hep-th/0006235.

\bibitem{fur} K. Furuta and T. Inami: ``Ultraviolet property of noncommutative
Wess-Zumino-Witten model", hep-th/0004024.

\bibitem{boss} M. Boss, Ann. of Phys. {\bf 181} (1988), 177.

\bibitem{zinn} J. Zinn-Justin:``Quantum field theory and critical phenomena",
Clarendon Press, Oxford (1993).

\bibitem{pes} M. Peskin and D. Schroeder: ``An introduction to quantum field theory",
Addison-Wesley Publishing Co., New York (1995).

\bibitem{goddol} P. Goddard and D. Olive, Int. Journ. Mod. Phys. {\bf A1} (1986), 303.

\bibitem{fiduno} E. Moreno and F. Schaposnik, ``The Wess-Zumino-Witten term in noncommutative
two dimensional fermion models", hep-th/0002236.

\bibitem{fiddos} E. Moreno and F. Schaposnik, ``Wess-Zumino-Witten and fermion models in
noncommutative space", hep-th/0008118.

\bibitem{gomis3} Ja. Gomis and T. Mehen: ``Space-time non-commutative field theories and
unitarity", hep-th/0005129.

\bibitem{peter} B. de Wit, M. Grisaru and P. van Nieuwenhuizen,
Nucl. Phys. B{\bf 408} (1993), 299.

\bibitem{leut} H. Leutwyler and M. Shifman, Int. J. Mod. Phys A{\bf 7} (1992), 795.

\bibitem{brezinn} E. Br\'{e}zin, J. Zinn-Justin and J. Le Guillou,
Phys. Rev D{\bf 14} (1976), 2615.

\bibitem{lagfm} L. Alvarez-Gaum\'{e}, D. Freedman and S. Mukhi, Ann. of Phys. 134 (1981), 85.

\bibitem{hull} C. Hull and P. Townsend, Nucl. Phys. B{\bf 274} (1986), 349.

\bibitem{grad} I. Gradshteyn and I. Ryzhik: ``Table of integrals, series and products",
Academic Press, New York (1965).


\end{thebibliography}
\end{document}